# Long memory, fractional integration and cointegration analysis of real convergence in Spain*

**Mariam Kamal[1] and Josu Arteche[2]**

*University of the Basque Country*

March 23, 2023.

**ABSTRACT**

This paper investigates economic convergence in terms of real income per capita among the autonomous regions of Spain over the period 1955-2020. In order to converge, the series should cointegrate. This necessary condition is checked using two testing strategies recently proposed for fractional cointegration, finding no evidence of cointegration, which rules out the possibility of convergence between all or some of the Spanish regions. As an additional contribution, an extension of the critical values of Nielsen's (2010) test of fractional cointegration is provided for a different number of variables and sample sizes from those originally provided by the author, fitting those considered in this paper.

*Corresponding address: Department of Applied Economics III (Econometrics and Statistics) at the University of the Basque Country, UPV-EHU, Bilbao 48015, Spain. Tel.: (+34) 946.013.740. *E-mail addresses: mariam.kamal@ehu.eus (permanent:mariam.kamal.economics@gmail.com)[1] and josu.arteche@ehu.eus.[2]

The authors acknowledge financial support from the grant[1] PRE_2018_1_0088 awarded by the Department of Education, Linguistic Policy and Culture of the Basque Government, the Spanish Ministry of Science and Innovation grant[2] PID2019-105183GB-I00 funded by MCIN/AEI/ 10.13039/501100011033 and UPV/EHU Econometrics Research[2] Group (Basque Government grant IT1359-19).

*Mariam wishes to express her heartfelt gratitude to an anonymous researcher for his extremely valuable and profound help.*

# 1. Introduction

Economic convergence has been one of the main focal points of the empirical literature on economic growth. It implies that income gaps between countries/regions tends to disappear, hence signifying convergence to a single steady state (equilibrium).

The European economy has become more integrated in recent decades, with the states following a converging path due to economic, political, and institutional factors, such as, for example, the exchange rate mechanism in 1979, and the introduction of the euro in 2001. Numerous empirical studies have shown this integration in the European Union (Caporaso and Pelowski, 1971, Martin and Ross 2004, Beckfield, 2006). However, this integration among states can come together with economic disparities between the different regions, which may cause non-convergence within that country. This paper analyses this possibility in Spain.

Spain's regions are divided into 17 Autonomous Communities. Some of these regions have more income than others due to their economic or sector specialization and disaggregation according to branches of activity. The income of each autonomous community depends on the economic specialization of that region, with some specializations generating low incomes while others generate high incomes. According to these criteria, the regions of Spain can be divided into two groups: (i) economically developed Spain, which are: Balearic Islands, Catalonia, Navarra, Madrid, the Basque Country, La Rioja, Aragón and the Valencian Community, which have a greater number of productive branches and present more specialization in branches with greater productivity, for example transport and telecommunications, chemicals, textiles, footwear, mechanical transformation, paper and printing; and (ii) a group of less developed regions: Extremadura, Andalucía, Galicia, Castilla-La Mancha and Castilla y León, whose wealth comes mainly from the primary sector — construction and food.

The different ways of obtaining income among different regions have generated the current regional differences in productivity and income. Hence, it might be unsurprising to find disparities between the 17 Autonomous Communities.

This prompts us to question whether economic convergence among the 17 Autonomous Communities could not occur, which is a question that drives the current research. In addition, we also aim to determine whether some converging subgroups can be identified, for example among developed or less developed regions.

As a first example of this heterogeneity across the Spanish regions, Figure 1 displays the evolution of the cross-sectional standard deviations for all the log of per capita incomes in the 17 autonomous communities in Spain from 1955 until 2020. The dispersion begins in 1955 at around 0.91 and rises and declines over time ending at around 0.94 in 2020, confirming the absence of sigma convergence.

Figure 1. Cross sectional standard deviation of the log of per capita income

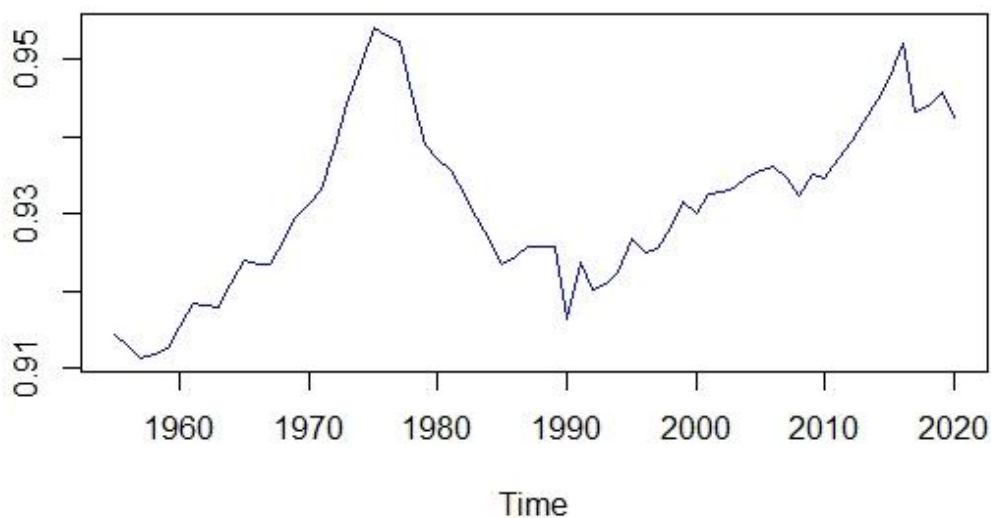

The large dispersion shown in the graph confirms our previous suspicion of possible heterogeneity and non-convergence. Note that the dispersions in the decade of 1970 and 2010 are the highest. However, the heterogeneity in Spain remains in the sample before and after these peaks, which may hinder regional convergence.

In the literature on economic growth, there are three main definitions of convergence: (i) beta convergence, (ii) sigma convergence and (iii) stochastic convergence. As the first two have several statistical problems (see Friedman, 1992, Quah, 1993 and Durlauf 2000), we will focus on the time-series approach on cointegration, as suggested by Bernard and Durlauf (1995). The use of different techniques has usually led to differing conclusions on the existence of convergence (see Durlauf, 2000 who find disparities in cross-sectional techniques). We follow a time-series approach to test for output convergence, paying particular attention to the analysis of cointegration, which provides a natural setting for testing relations between variables (see Quah, 1993, 1994, Evans, 1995, Bernard and Durlauf, 1995, 1996 and Durlauf 2000).

According to the time series approach of Bernard and Durlauf (1995, 1996), two series converge if the following conditions are satisfied: (i) The variables are cointegrated, (ii) the cointegrating vector is (1,-1), and (iii) The difference between the series is a stochastic variable with zero mean. Based on these conditions, the notion of convergence can be divided into strong and weak convergence (defined as catching-up in the convergence literature). If conditions (i) and (ii) are fulfilled, the series are cointegrated with cointegrating vector $[\mathbf{1}, -\mathbf{1}]$, but the difference between them is a stochastic variable with a mean different from zero, which suggests that the deviation between the series is expected to decrease, but not disappear. This is weak convergence, i.e., catching up, which refers to the situation in which the narrowing of the differences between the variables is observed over time, but the convergence process has yet to be complete. If all conditions (i), (ii), and (iii) are fulfilled there is strong convergence in the long run because the difference between the variables vanishes. Therefore, if there is no cointegration, convergence does not occur, neither weak nor strong. This stochastic convergence examines whether permanent movements in one country/region per capita income are related to permanent movements in another country/region per capita income.

This paper analyses convergence in annual real output per capita of the 17 autonomous communities in Spain from 1995 to 2020. We contribute to the existing empirical literature in two main dimensions:

- First, we examine real income convergence among the income of Spanish regions using novel semiparametric and nonparametric techniques proposed to test for fractional cointegration, in particular the techniques proposed by Robinson (2008), Hualde (2012) and Nielsen (2010). These fractional integration and cointegration techniques are used to avoid the low power of traditional unit root and cointegration tests against fractional alternatives.
- Second, Nielsen (2010) only provides critical values for 8 variables and a sample size of 1000 observations for the application of his testing strategy. However, our data set comprises 17 variables and 66 observations, so as an additional contribution, the critical values are here extended to a greater number of variables and different sample sizes T=66, T=150 and T=1000.

The rest of the paper is organized as follows. *Section 2* explains the methodologies used in our analysis. *Section 3* contains the data and presents the results, and finally, *Section 4* presents the Discussion and Conclusions.

## 2. Methodology

The methodology used in this paper is based on the concepts of fractional integration and fractional cointegration

### 2.1. Fractional integration and cointegration

The idea of fractional integration was introduced by Granger and Joyeux (1980), Granger (1980, 1981) and Hosking (1981) allowing a continuous transition from non-unit to unit root behaviours, offering a more flexible context for the modelling of long-run persistence. A time-series $\{y_t, t = 1,2,3, \dots \}$ is (fractionally) integrated of order d, $I(d)$, if it satisfies:

$$(1 - L)^d y_t = u_t, \quad t = 0, \pm 1, , \dots, \qquad (1)$$

where d is the memory parameter and $u_t \sim I(0)$, meaning that $u_t$ has a finite variance and a spectral density function $f(w)$, satisfying $0 < f(w) < \infty$. If $d = 0$, $y_t = u_t$ and $y_t$ is said to be short memory; if $0 < d < \frac{1}{2}$, $y_t$ is said to be long memory. Finally, if $-\frac{1}{2} < d < 0$, $y_t$ presents anti-persistence. Also, if $d < 0.5$, $y_t$ is covariance stationary. However, a value $d \geq 0.5$ implies non-stationarity, but if $d < 1$, the series is mean-reverting. In addition, if $d = 1$, the series has a unit root. The difference between $d < 1$ and $d \geq 1$ is that if $d < 1$ the effects of the shocks disappear in the long-run and if $d \geq 1$ the shocks persist indefinitely.

Note that $u_t$ in (1) may include some type of weak dependence in the form of, for example, a stationary and invertible autoregressive moving average (ARMA) process:

$$\Phi(L)u_t = \theta(L)\varepsilon_t, \ \ t = 0, \pm 1, \ldots, \tag{2}$$

where $\varepsilon_t$ is an independent and identically distributed (*iid*) sequence. In this case, $y_t$ in (1) is an Autoregressive Fractionally Integrated Moving Average (ARFIMA) process:

$$\Phi(L)(1-L)^d y_t = \theta(L)\varepsilon_t, \ \ t = 0, \pm 1, , \ldots \tag{3}$$

Several parametric and semiparametric techniques currently exist for the estimation of d, the gaussian semiparametric by Robinson (1995a), log-periodogram by Robinson (1995b), exact local Whittle by Shimotsu and Phillips (2005) and exact local Whittle with unknown mean and time trend by Shimotsu (2010), among others).

Engle and Granger (1987) defined cointegration as follows: "*A vector $y_t$ is said to be co-integrated of order d, b, denoted $y_t \sim CI(d, b)$, if the components of $y_t$ are $I(d)$ and there exists a vector $\alpha(\neq 0)$ such that $z_t = \alpha' y_t \sim I(d-b), \ b > 0$. The vector α is called the co-integrating vector and b denotes the degree of cointegration.*

The original testing strategies proposed for cointegration were only suitable for bivariate settings and they thus could only identify one cointegration vector. Johansen (1988, 1991, 1995) developed a maximum likelihood approach for testing cointegration in a multivariate setting allowing for several relations and determining the rank of cointegration. Following these pioneering authors, other standard techniques were developed by Harris (1997), Bierens (1997), Phillips and Ouliaris (1988) and Breitung (2002), among others. The generalization of the traditional Johansen test is the fractionally cointegrated vector autoregressive (FCVAR) model in a fractional context which was proposed by Johansen (2008) and Johansen and Nielsen (2010, 2012, 2014). Standard traditional cointegration is just one particular case of fractional cointegration where the memory parameters $d$ and the degree of cointegration $b$ are restricted to be integer values. Fractional values of $d$ and $b$ allow for more flexibility and are good alternatives because many economic series are known to exhibit non-stationary behaviours that may not be exactly I(1), and also there is no need to assume that the unobservable equilibrium relation is exactly I(0).

## 2.2. Testing for fractional cointegration

The strategy we follow is based on the estimation of the cointegration rank in a fractional setting using two different and flexible techniques with good asymptotic properties under mild conditions. First, the methodology proposed by Nielsen (2010) has the following advantages over other cointegration tests: (i) the test statistic is calculated without prior knowledge of the order of integration of the series. (ii) Since the test is nonparametric, it does not require specification of a particular model and is invariant to short-run dynamics. This is important because mis-specified short-run dynamics may lead to inconsistent estimation and hence to erroneous inference regarding the cointegration rank in other parametric techniques. (iii) The proposed test has good power for large and small samples. Second, the methodology offered by Hualde (2012), together with the testing strategy used by Robinson (2008), is characterized by the following: (i) The testing strategies in Robinson (2008) do not require estimation of any cointegrating relations or tuning numbers beyond one bandwidth parameter. (ii) Hualde (2012) proposes a procedure to estimate the rank of cointegration in multivariate fractional series, and therefore this can be implemented together with the procedure in Robinson (2008) to infer the dimension of the possible cointegrating subspaces. (iii) The combination of both techniques allows for precise detection of the common trends.

**Robinson (2008) and Hualde's (2012) approaches**

We first consider the test proposed by Robinson (2008) combined with the strategy in Hualde (2012). Hualde's procedure has the advantage over other fractional cointegration approaches of providing an automatic method for inferring cointegrating relationships without any prior information about the variables. He defines the possibility of cointegration as a situation in which a linear combination of fractional processes is integrated in a strictly smaller order than the maximum order of the elements of the linear combination. For example, if one of the variables has an integration order strictly larger than the rest of the variables, any linear combination which puts zero weight on this particular variable is considered to be a trivial cointegrating relation. Under this definition, the variables can have different integration orders. However, when all the variables have the same integration order, this definition coincides with that originally provided by Engle and Granger (1987).

The proposal by Hualde (2012) is based on an estimator of the cointegrating rank, $r$, obtained by applying sequentially the procedure discussed in his Theorem 1, which we rewrite here:

**Theorem 1** (Hualde, 2012). $y_t$ has cointegrating rank $r \in \{1, ..., p-1\}$ where $p$ is the number of variables in the vector $y_t$, if (i) and (ii) are satisfied, where: (i) There exists a $(p-r)$ dimensional subvector of $y_t$, denoted as $y_{(b)t}$, whose individual components are common trends, denoted as CT. (ii) All subvectors of $y_t$ of dimension larger than $p-r$ containing $y_{(b)t}$ cointegrate.

The procedure to estimate the rank of cointegration r is based on the following steps. First, the estimates of the integration orders (memory parameters), $\hat{d}_i$, $i = 1, ..., p$ are obtained to define the CT as the series with the highest order of integration. Then, the following hypothesis is tested:

$$H_{j_1,...,j_k}: \{y_{j_1 t}, y_{j_2 t}, ..., y_{j_k t} \text{ are not cointegrated}\};$$

against

$$\overline{H}_{j_1,...,j_k}: H_{j_1,...,j_k} \text{ is not true,}$$

where $j_1 ..., j_k \in \{1, ..., p\}$, $k \leq p$, is sequentially tested. In order to estimate the memory parameters, the univariate local Whittle estimator $\hat{d}_i$, proposed by Robinson (1995a) is used.

Next, the hypotheses are tested using the statistic X* proposed by Robinson (2008), defined as:

$$X^* = m s^*(\tilde{d})^2 / \{p^2 \, tr(\hat{R}^* A \hat{R}^* A) - p\} \qquad (4)$$

where

$m$ is the bandwidth

$$s^*(\tilde{d}) = tr\{\hat{G}^*(\tilde{d})^{-1} \hat{H}^*(\tilde{d})\}$$

$$\hat{G}^*(d) = \frac{1}{m} \sum_{j=1}^{m} I_y(\lambda_j) \lambda_j^{2d}$$

$$\hat{H}^*(d) = \frac{1}{m} \sum_{j=1}^{m} v_j I_y(\lambda_j) \lambda_j^{2d}$$

$$v_j = \log j - \frac{1}{m}\sum_{i=1}^{m} \log i$$

$$\hat{R}^* = \hat{D}^{1/2}\hat{G}^*(\tilde{d})\,\hat{D}^{-1/2}$$

$$\hat{D} = diag\{\hat{g}_{11}, \dots, \hat{g}_{pp}\}, \text{ where } \hat{g}_{ii} \text{ is the ith diagonal element of } \hat{G}^*(\tilde{d})$$

$$A = diag\{a_1, \dots, a_p\}$$

where $\tilde{d} = \sum_{i=1}^{p} a_i \hat{d}_i$ and the $a_i$ are arbitrarily chosen weights satisfying that $\sum_{i=1}^{p} a_i = 1$. For instance, Robinson (2008) takes $a_i \equiv 1/p$, so the arithmetic mean of the $\hat{d}_i$ is used. Another option is using $a_j = 1$, $a_i = 0, i \neq j$ some j. In our case, we use the first option as recommended by Robinson (2008).

Under the null hypothesis of non-cointegration and stationarity of the series, which implies that all the memory parameters are smaller than 0.5,

$$X^* \xrightarrow{d} X_1^2 \quad as\ T \to \infty. \tag{5}$$

The methodology to estimate the cointegration rank r is characterized by the following steps:

**Step 1.** Estimate the individual integration orders, $d_i$, by $\hat{d}_i$, $i = 1, \dots, p$. Then choose a possible CT $y_{c_1 t}$ as the variable with the highest estimated order, such that $c_1 \in \{1, \dots, p\}$. Next, reorder the variables in $y_t$ so that $y_{pt} = y_{c_1 t}$ in the new ordering. Finally, given the possible CT i.e., $y_{pt}$, we test the following hypotheses:

$H(1): \cup_{i=1}^{p-1} H_{p,i}$ versus $\bar{H}(1): \cap_{i=1}^{p-1} \bar{H}_{p,i}$. Note that $H(1)$ means non-cointegration in pairs of each variable with the CT, and $\bar{H}(1)$ means that $H(1)$ is not true. The process ends if $H(1)$ is rejected, and so it is concluded that $\hat{r} = p - 1$. Otherwise, it is not rejected, and the process continues to Step 2. Consequently, following Theorem 1, the hypotheses are equivalent to $r < p - 1$ and $r = p - 1$ respectively.

**Step 2.** If $H(1)$ is not rejected, choose a second possible CT as the variable with the smallest statistic $X^*$ i.e., $y_{c_2 t}, c_2 \in \{1, \dots, p-1\}$. There will be two possible CTs altogether. These CTs are denoted as $y_{pt}$ and $y_{c_2 t}$ respectively. Then we reorder again the variables so that $y_{pt} = y_{c_1 t}$ and $y_{p-1,t} = y_{c_2 t}$ in the new ordering. Finally, given the possible CTs i.e., $y_{pt}, y_{p-1,t}$ we test the following hypotheses: $H(2): \cup_{i=1}^{p-2} H_{p,p-1,i}$ versus $\bar{H}(2): \cap_{i=1}^{p-2} \bar{H}_{p,p-1,i}$. Note that $H(2)$ means non-cointegration for any set of three variables containing the CTs $y_{pt}, y_{p-1,t}$, and $\overline{H}(2)$ means that H(2) is not true. Then, the hypotheses are equivalent to $r < p - 2$ and $r = p - 2$ respectively and the process ends if H(2) is rejected.

**Step $k$** ($for\ k = 2, \ldots, p - 1$). If $H(k - 1)$ is not rejected, choose $c_k$. Sort the variables so that $y_{pt} = y_{c_1 t}, \ldots, y_{p-k+2,t} = y_{c_{k-1},t}$ and choose the possible CTs, as previously.

Finally, test the following hypothesis:

H(k): $\cup_{i=1}^{p-k} H_{p,p-1,\ldots,p-k+1,i}$ versus $\overline{H}(k): \cap_{i=1}^{p-k} \overline{H}_{p,p-1,\ldots,p-k,i}$ and if there is cointegration, the estimation will be $\hat{r} = p - k$. However, if we reach the last step $k = p - 1$ this means that $\hat{r} = 0$ and $H(i), i = 1, 2, 3, \ldots, p - 1,$ are not rejected.

The testing procedure based on the statistic X* has low power with small sample sizes, which can significantly influence the results obtained when analysing the possibility of cointegration in the Spanish regions. To complement the results obtained we also consider the test proposed by Nielsen (2010), which has greater power for small samples, (*see Nielsen's Monte Carlo*).

**Nielsen's (2010) approach and critical values extension**

The test statistic is defined as follows:

$$\Lambda_{p,r}(d_1) = T^{2d_1} \sum_{j=1}^{p-r} \vartheta_j, \qquad r = 0, \ldots, p - 1 \qquad (6)$$

where $\vartheta_j, j = 1, \ldots, p,$ are the eigenvalues of $|\vartheta B_T - A_T| = 0$, for $A_T = \sum_{t=1}^{T} Z_t Z_t'$, $B_T = \sum_{t=1}^{T} \tilde{Z}_t \tilde{Z}_t'$, $\tilde{Z}_t = \Delta_1^{-d_1} Z_t$ with $d_1 > 0$, $t = 1, 2, \ldots, T$, and $Z_t$ is the p-vector of time series under analysis (perhaps after extracting deterministic terms), which is fractionally integrated of order d, where d is a vector containing the individual orders of integration of the elements in $Z_t$, which possibly differ from each other. Note that (6) defines a family of tests indexed by the fractional integration parameter, $d_1$. Nielsen (2010) argues in favour of using $d_1$=0.1 based on an asymptotic local power analysis and on simulations. For this reason, we use this value in the empirical application. Large values of $\Lambda_{p,r_0}(d_1)$ are associated with the rejection of the null hypothesis $H_0: r = r_0$ versus $H_1: r > r_0$. Nielsen's procedure has the advantage of not requiring knowledge of the fractional integration and cointegration orders *d* and *b* as long as the series are non-stationary, implying memory parameters greater than 0.5. However, its asymptotic distribution is non-standard, but Nielsen (2010) simulated critical values for p<8 variables with a sample size of 1000 to facilitate its application. We complement Nielsen's (2010) tables by providing more critical values to cover up to 17 variables for all models.

The observed time series $\{Y_t\}_{t=1}^T$ considered by Nielsen (2010) is generated by

$$Y_t = \alpha'\delta_t + Z_t, \qquad t = 1,2,\ldots \qquad (7)$$

where $\delta_t$ may contain deterministic terms. Three different cases are analysed: $\delta_t = 0$ when there are no deterministic terms, $\delta_t = 1$ when there is a non-zero mean, and $\delta_t = [1,t]'$ when there is a deterministic trend. The critical values in Nielsen (2010) in his Tables 1, 2 and 3 are here extended for these three cases for samples T=1000, 150 and 66, in Tables 1-3, where the latter is the number of observations used in our application. All tables are based on 100000 replications.

Table 1. Simulated critical values $CV_{\xi,p-r}(1,d_1)$ $for\ large\ sample$ $(T=1000)$ with $\delta_t=0$, $\delta_t=1$ and $\delta_t=[1,t]'$

| $d_1$ | $\xi$ | $p-r$ | | | | | | | | | | | | | | | | |
|---|---|---|---|---|---|---|---|---|---|---|---|---|---|---|---|---|---|---|
| | | 1 | 2 | 3 | 4 | 5 | 6 | 7 | 8 | 9 | 10 | 11 | 12 | 13 | 14 | 15 | 16 | 17 |
| | | $CV_{\xi,p-r}(1,d_1)$ $for$ $\delta_t=0$ | | | | | | | | | | | | | | | | | |
| 0.1 | 0.10 | 1.54 | 3.07 | 4.78 | 6.60 | 8.51 | 10.49 | 12.54 | 14.64 | 16.79 | 19.00 | 21.24 | 23.53 | 25.85 | 28.21 | 39.60 | 33.02 | 35.47 |
| | 0.05 | 1.62 | 3.16 | 4.86 | 6.68 | 8.59 | 10.57 | 12.52 | 14.73 | 16.89 | 19.09 | 21.33 | 23.62 | 25.94 | 28.30 | 30.69 | 33.12 | 35.57 |
| | 0.01 | 1.77 | 3.33 | 5.03 | 6.85 | 8.75 | 10.74 | 12.80 | 14.90 | 17.06 | 19.26 | 21.50 | 23.79 | 26.11 | 28.48 | 30.87 | 33.30 | 35.75 |
| | | $CV_{\xi,p-r}(1,d_1)$ $for$ $\delta_t=1$ | | | | | | | | | | | | | | | | | |
| | 0.10 | 1.76 | 3.50 | 5.32 | 7.23 | 9.21 | 11.26 | 13.36 | 15.52 | 17.71 | 19.95 | 23.61 | 24.56 | 26.92 | 29.31 | 31.73 | 34.18 | 36.65 |
| | 0.05 | 1.82 | 3.57 | 5.40 | 7.31 | 9.29 | 11.34 | 13.45 | 15.60 | 17.80 | 20.04 | 23.68 | 24.65 | 27.01 | 29.40 | 31.82 | 34.27 | 36.75 |
| | 0.01 | 1.94 | 3.71 | 5.54 | 7.46 | 9.45 | 11.50 | 13.61 | 15.76 | 17.97 | 20.21 | 23.80 | 24.82 | 27.18 | 29.57 | 31.99 | 34.45 | 36.92 |
| | | $CV_{\xi,p-r}(1,d_1)$ $for$ $\delta_t=[1,t]'$ | | | | | | | | | | | | | | | | | |
| | 0.10 | 1.93 | 3.81 | 5.75 | 7.74 | 9.80 | 11.90 | 14.06 | 16.26 | 18.50 | 20.79 | 23.11 | 25.47 | 27.85 | 30.28 | 32.72 | 35.20 | 37.71 |
| | 0.05 | 1.98 | 3.88 | 5.82 | 7.82 | 9.88 | 11.99 | 14.15 | 16.35 | 18.50 | 20.87 | 23.20 | 25.55 | 27.94 | 30.37 | 32.82 | 35.29 | 37.80 |
| | 0.01 | 2.08 | 4.01 | 5.97 | 7.97 | 10.04 | 12.15 | 14.31 | 16.51 | 18.76 | 21.04 | 23.36 | 25.72 | 28.12 | 30.54 | 32.99 | 35.47 | 37.98 |

Note: The simulated critical values are based on 100000 replications for up to 17 series.

Table 2. Simulated critical values $CV_{\xi,p-r}(1,d_1)$ for small samples (T=66) with $\delta_t= 0$, $\delta_t= 1$ and $\delta_t= [1, t]$'

| $d_1$ | $\xi$ | \multicolumn{17}{c}{$p-r$} |
|---|---|---|---|---|---|---|---|---|---|---|---|---|---|---|---|---|---|---|
|   |   | 1 | 2 | 3 | 4 | 5 | 6 | 7 | 8 | 9 | 10 | 11 | 12 | 13 | 14 | 15 | 16 | 17 |
|   |   | \multicolumn{17}{c}{$CV_{\xi,p-r}(1,d_1)$ for $\delta_t = 0$} |
| 0.1 | 0.10 | 1.52 | 3.03 | 4.71 | 6.49 | 8.35 | 10.27 | 12.25 | 14.27 | 16.33 | 18.43 | 20.56 | 22.71 | 24.90 | 27.11 | 29.34 | 31.58 | 33.85 |
|     | 0.05 | 1.60 | 3.11 | 4.79 | 6.52 | 8.42 | 10.34 | 12.32 | 14.34 | 16.40 | 18.50 | 20.62 | 22.78 | 24.97 | 27.17 | 29.40 | 31.65 | 33.91 |
|     | 0.01 | 1.74 | 3.28 | 4.94 | 6.72 | 8.57 | 10.49 | 12.46 | 14.48 | 16.53 | 18.63 | 20.76 | 22.91 | 25.09 | 27.29 | 29.52 | 31.77 | 34.03 |
|   |   | \multicolumn{17}{c}{$CV_{\xi,p-r}(1,d_1)$ for $\delta_t = 1$} |
|     | 0.10 | 1.74 | 3.44 | 5.23 | 7.10 | 9.02 | 11.00 | 13.03 | 15.09 | 17.19 | 19.33 | 21.49 | 23.67 | 25.88 | 28.12 | 30.37 | 32.64 | 34.92 |
|     | 0.05 | 1.79 | 3.51 | 5.30 | 7.17 | 9.09 | 11.07 | 13.10 | 15.16 | 17.26 | 19.39 | 21.55 | 23.74 | 25.95 | 28.18 | 30.43 | 32.70 | 34.98 |
|     | 0.01 | 1.90 | 3.64 | 5.44 | 7.30 | 9.23 | 11.20 | 13.22 | 15.29 | 17.39 | 19.59 | 21.68 | 23.86 | 26.07 | 28.29 | 30.54 | 32.81 | 35.09 |
|   |   | \multicolumn{17}{c}{$CV_{\xi,p-r}(1,d_1)$ for $\delta_t = [1,t]$'} |
|     | 0.10 | 1.89 | 3.73 | 5.63 | 7.57 | 9.55 | 11.58 | 13.65 | 15.75 | 17.88 | 20.05 | 22.23 | 24.44 | 26.68 | 28.93 | 31.20 | 33.48 | 35.79 |
|     | 0.05 | 1.94 | 3.79 | 5.69 | 7.63 | 9.62 | 11.65 | 13.72 | 15.82 | 17.95 | 20.11 | 22.30 | 24.51 | 26.74 | 28.99 | 31.26 | 33.54 | 35.84 |
|     | 0.01 | 2.02 | 3.90 | 5.81 | 7.76 | 9.75 | 11.78 | 13.85 | 15.95 | 18.08 | 20.24 | 22.42 | 24.63 | 26.86 | 29.10 | 31.37 | 33.66 | 35.96 |

Note: The simulated critical values are based on 100000 replications and sample size 66 for up to 17 series.

Table 3. Simulated critical values $CV_{\xi,p-r}(1,d_1)$ for small samples size (T=150) with $\delta_t= 0$, $\delta_t= 1$ and $\delta_t= [1, t]$'

| $d_1$ | $\xi$ | \multicolumn{17}{c}{$p-r$} |
|---|---|---|---|---|---|---|---|---|---|---|---|---|---|---|---|---|---|---|
| | | 1 | 2 | 3 | 4 | 5 | 6 | 7 | 8 | 9 | 10 | 11 | 12 | 13 | 14 | 15 | 16 | 17 |
| | | \multicolumn{17}{c}{$CV_{\xi,p-r}(1,d_1)$ for $\delta_t = 0$} |
| 0.1 | 0.10 | 1.53 | 3.05 | 4.75 | 6.55 | 8.44 | 10.40 | 12.42 | 14.49 | 16.61 | 18.77 | 20.97 | 23.21 | 25.47 | 27.77 | 30.09 | 32.44 | 34.81 |
| | 0.05 | 1.61 | 3.14 | 4.83 | 6.64 | 8.52 | 10.48 | 12.50 | 14.58 | 16.69 | 18.85 | 21.05 | 23.28 | 25.55 | 27.85 | 30.17 | 32.52 | 34.89 |
| | 0.01 | 1.76 | 3.31 | 4.99 | 6.79 | 8.68 | 10.64 | 12.66 | 14.74 | 16.85 | 19.01 | 21.21 | 23.44 | 25.71 | 28.00 | 30.32 | 32.67 | 35.04 |
| | | \multicolumn{17}{c}{$CV_{\xi,p-r}(1,d_1)$ for $\delta_t = 1$} |
| | 0.10 | 1.75 | 3.47 | 5.28 | 7.18 | 9.14 | 11.16 | 13.23 | 15.35 | 17.51 | 19.71 | 21.94 | 24.21 | 26.50 | 28.83 | 31.18 | 33.56 | 35.95 |
| | 0.05 | 1.81 | 3.54 | 5.36 | 7.25 | 9.21 | 11.24 | 13.31 | 15.43 | 17.59 | 19.79 | 22.02 | 24.29 | 26.58 | 28.91 | 31.26 | 33.63 | 36.03 |
| | 0.01 | 1.92 | 3.68 | 5.50 | 7.40 | 9.36 | 11.38 | 13.46 | 15.58 | 17.74 | 19.94 | 22.18 | 24.44 | 26.73 | 29.06 | 31.41 | 33.78 | 36.17 |
| | | \multicolumn{17}{c}{$CV_{\xi,p-r}(1,d_1)$ for $\delta_t= [1, t]$'} |
| | 0.10 | 1.91 | 3.78 | 5.70 | 7.67 | 9.70 | 11.78 | 13.90 | 16.06 | 18.26 | 20.50 | 22.76 | 25.06 | 27.38 | 29.74 | 32.11 | 34.50 | 36.92 |
| | 0.05 | 1.96 | 3.84 | 5.77 | 7.75 | 9.77 | 11.85 | 13.98 | 16.14 | 18.34 | 20.57 | 22.84 | 25.14 | 27.46 | 29.81 | 32.18 | 34.58 | 37.00 |
| | 0.01 | 2.05 | 3.97 | 5.90 | 7.89 | 9.92 | 12.00 | 14.12 | 16.29 | 18.49 | 20.72 | 22.99 | 25.29 | 27.61 | 29.96 | 32.33 | 34.73 | 37.15 |

Note: The simulated critical values are based on 100000 replications and sample size 150 for up to 17 series.

## 3. Empirical analysis

### 3.1. Preliminary analysis of the variables

The data used are the logarithms of the annual real GDP (constant 2010 €) of the 17 autonomous communities in Spain (omitting the autonomous cities) in thousands of euros from 1955 to 2020 for a total of T=66 observations. In order to apply the semiparametric cointegration analysis by Robinson (2008), the series are differenced to obtain growth rates that are stationary, whereas for the nonparametric cointegration analysis by Nielsen (2010), the series are raw series (non-stationary). The variables in logarithms are denoted by the name of the autonomous community, and the growth rate is denoted by the abbreviation of these autonomous communities (in parenthesis the notation of the growth rates): Andalucia (andal), Aragon (ara), Asturias (ast), Balearic Islands (bal), Canary Islands (can), Cantabria (cant), Catalonia (cat), Castilla-LaMancha (clm), Castilla-Leon (cyl), Extremadura (ext), Galicia (gal), Madrid (mad), Murcia (mur), Navarre (nav), Basque Country (pv), Rioja (rio) and Valencia (val). All data were provided by FEDEA (Foundation for the Study of Applied Economics) and INE (Spanish National Statistics Institute).

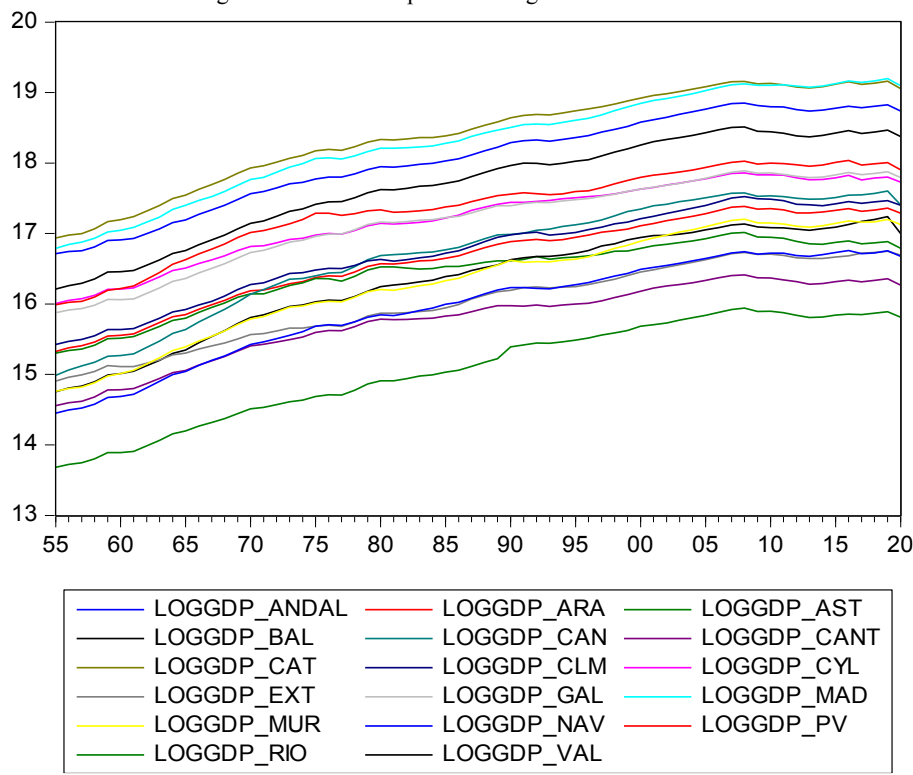

Figure 2. Time series plot of all log real GDP variables

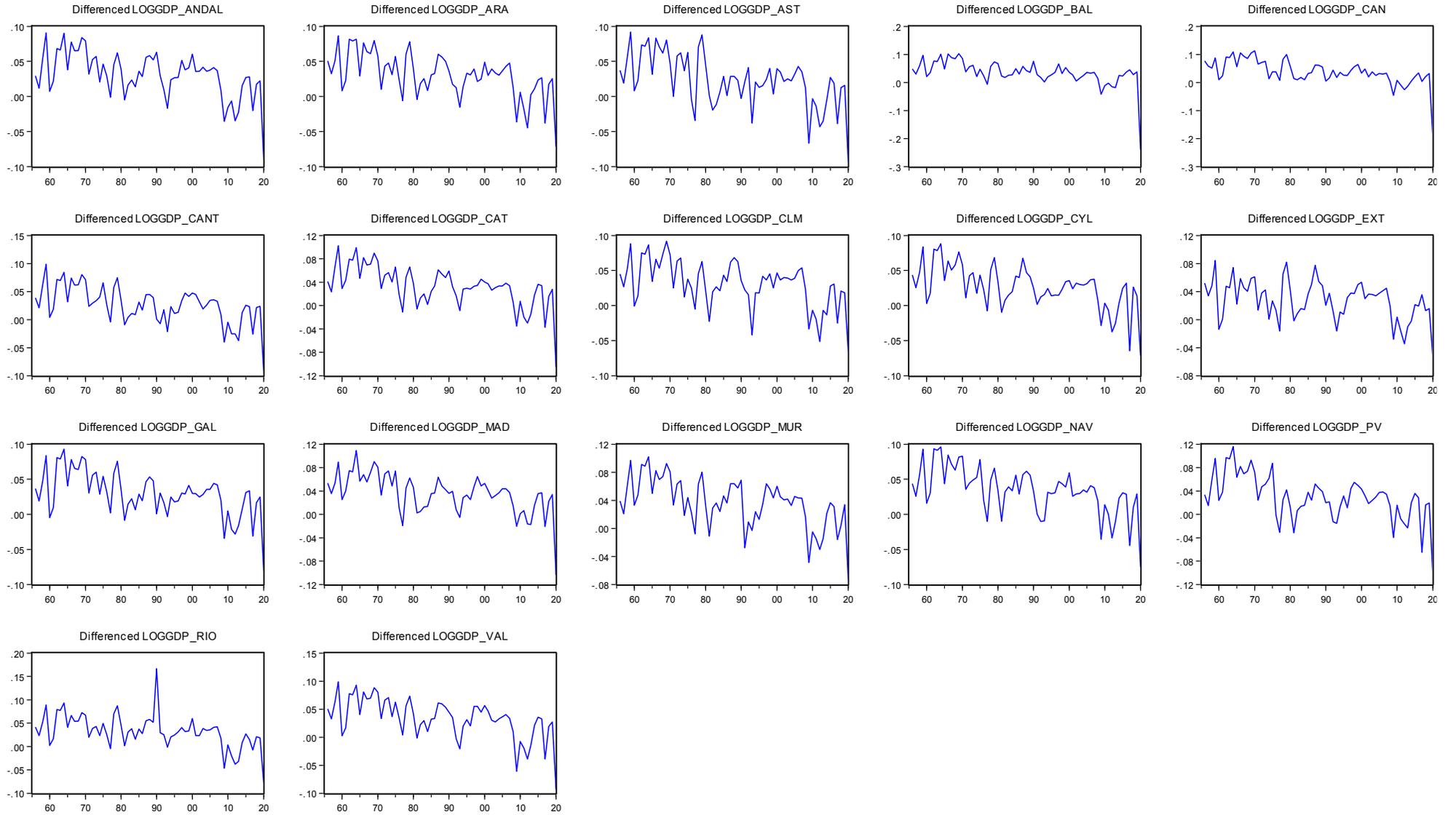

Figure 3. Time series plots of the GDP growth rate

In order to shed more light on the persistence of the series, the Exact Local Whittle (ELW) estimator proposed by Shimotsu and Phillips (2005), which is consistent and asymptotically normal for any value of $d$, was applied on the raw series. All the estimates of the memory parameters (available on request) are between 1 and 1.4, confirming the non-stationarity of the series, which is a requirement for the applicability of Nielsen's procedure. The estimation of the memory parameter of the growth rates, required for the application of Robinson's (2008) test is, however, obtained using the Local Whittle estimator of Robinson (1995b) as suggested in that paper, which is consistent for d<1 and asymptotically normal for d<0.75, obtaining estimates between 0 and 0.5 (see Table 5).

### 3.2. Robinson (2008) and Hualde (2012) cointegration results

In order to analyse the robustness of the results to the selection of the bandwidth, the entire analysis has been implemented using four different bandwidths, m=18, m=20, m=23 and m=28. Table 4 shows that the Local Whittle estimates of the first differenced logs are between 0.2 and 0.49, indicating that the growth rate series can be considered stationary ($d < 0.5$). According to the results of the estimated memory parameters, the possible CT in Step 1 (variable with the largest estimated d) is *Murcia (mur)* for all the bandwidths. In subsequent steps, i.e. 2, 3,…, p-1=16, the possible CTs are chosen as the series with the smallest statistic X* for each particular bandwidth.

To understand how Hualde's (2012) procedure works, we will describe in more detail some of its steps. Step 1 checks for cointegration, two by two. For instance, for a bandwidth $m = 18$, the results indicate that $mur_t$ is the possible common trend. Therefore, according to this choice of bandwidth, the first step in this procedure is to test:

$$H(1): H_{mur,andal} \cup H_{mur,ara} \cup H_{mur,ast} \cup H_{mur,bal} \cup H_{mur,cant} \cup H_{mur,cat} \cup H_{mur,clm} \cup H_{mur,cyl} \cup H_{mur,ext} \cup H_{mur,gal} \cup H_{mur,mad} \cup H_{mur,can} \cup H_{mur,nav} \cup H_{mur,pv} \cup H_{mur,rio} \cup H_{mur,val};$$

against

$$\bar{H}(1): \bar{H}_{mur,andal} \cap \bar{H}_{mur,ara} \cap \bar{H}_{mur,ast} \cap \bar{H}_{mur,bal} \cap \bar{H}_{mur,cant} \cap \bar{H}_{mur,cat} \cap \bar{H}_{mur,clm} \cap \bar{H}_{mur,cyl} \cap \bar{H}_{mur,ext} \cap \bar{H}_{mur,gal} \cap \bar{H}_{mur,mad} \cap \bar{H}_{mur,can} \cap \bar{H}_{mur,nav} \cap \bar{H}_{mur,pv} \cap \bar{H}_{mur,rio} \cap \bar{H}_{mur,val}$$

Where for

$$a_t = andal_t, ara_t, ast_t, bal_t, cant_t, cat_t, clm_t, cyl_t, ext_t, gal_t, mad_t, can_t, nav_t, pv_t, rio_t, val_t$$

(all variables without the CT)

$$H_{mur,a}: mur_t, a_t \text{ are not cointegrated}$$

$$\overline{H}_{mur,a}: H_{mur,a} \text{ is not true}$$

The above hypotheses are equivalent to $H(1): r < 16$ and $\overline{H}(1): r = 16$. According to Hualde's (2012) procedure, if $mur_t$ is a true CT, then r<16 if and only if H(1) holds, and r=16 if and only if $\overline{H}(1)$ holds.

The process ends when H(k) for $k = 1,2,...,16$ is rejected. In this case the process continues to Step 2 because the null hypothesis of no cointegration cannot be rejected for any pair of variables and for any bandwidth choice because none of the X* statistics are greater than the critical value of the $\chi_1^2$, that is, 3.84 at 5% level of significance (see Table 5). The second possible $CT$ is chosen as the series leading to the smallest X* statistics obtained in Step 1 (see also Table 5), that is $rio_t$ (La Rioja). The null hypothesis of no cointegration is not rejected in either of the steps. Full details can be found in Tables 8 to 21 in the ***supplementary material***. The first and last step are shown in Table 5 and Table 6, respectively. The last step tests if there is any long-run relationship between the 17 variables, i.e., the hypotheses are equivalent to $r < 1$ and $r = 1$ that is:

$$H(16): H_{can,cyl,pv,rio,cant,ext,gal,cat,clm,bal,andal,val,ast,ara,nav,mad,mur}$$

against

$$\overline{H}(16): \overline{H}_{can,cyl,pv,rio,cant,ext,gal,cat,clm,bal,andal,val,ast,ara,nav,mad,mur}$$

where

$$H_{can,cyl,pv,rio,cant,ext,gal,cat,clm,bal,andal,val,ast,ara,nav,mad,mur}:$$
$can_t, cyl_t, pv_t, rio_t, cant_t, ext_t, gal_t cat_t, clm_t, bal_t, andal_t, val_t, ast_t, ara_t, nav_t, mad_t, mur_t$ are not cointegrated

$$\overline{H}_{can,cyl,pv,rio,cant,ext,gal,cat,clm,bal,andal,val,ast,ara,nav,mad,mur}:$$
$H_{can,cyl,pv,rio,cant,ext,gal,cat,clm,bal,andal,val,ast,ara,nav,mad,mur}$ is not true

The results of the test in this step for various bandwidths are shown in Table 6. There is no evidence of cointegration for any of the bandwidths considered. Therefore, there is no long-run relationship between any of the 17 autonomous communities of Spain, implying that the growth rates of the output in these regions do not converge and they have some different structural characteristics across the autonomous communities. This result seems to run counter to the first impression obtained from Figures 2 and 3, which show a similar evolution, suggesting the existence of some long run equilibrium but reinforces the lack of sigma convergence observed in Figure 1. To further support this conclusion, the nonparametric method proposed by Nielsen (2010) was also used.

Table 4. Estimation of memory parameters at frequency 0

| Local Whittle | | | | |
|---|---|---|---|---|
| | m:18 | m:20 | m:23 | m:28 |
| **Andal** | 0.491 | 0.452 | 0.448 | 0.337 |
| **Ara** | 0.394 | 0.367 | 0.339 | 0.268 |
| **Ast** | 0.361 | 0.317 | 0.319 | 0.274 |
| **Bal** | 0.341 | 0.321 | 0.323 | 0.218 |
| **Can** | 0.432 | 0.375 | 0.332 | 0.259 |
| **Cant** | 0.397 | 0.368 | 0.378 | 0.310 |
| **Cat** | 0.448 | 0.453 | 0.410 | 0.326 |
| **Clm** | 0.496 | 0.422 | 0.383 | 0.323 |
| **Cyl** | 0.389 | 0.355 | 0.310 | 0.270 |
| **Ext** | 0.355 | 0.307 | 0.345 | 0.260 |
| **Gal** | 0.436 | 0.392 | 0.357 | 0.296 |
| **Mad** | 0.418 | 0.407 | 0.383 | 0.288 |
| **Mur** | 0.497 | 0.455 | 0.454 | 0.373 |
| **Nav** | 0.380 | 0.389 | 0.385 | 0.314 |
| **Pv** | 0.429 | 0.433 | 0.414 | 0.357 |
| **Rio** | 0.346 | 0.324 | 0.316 | 0.281 |
| **Val** | 0.447 | 0.404 | 0.387 | 0.311 |

Table 5. Cointegration rank test, step I

| **Step 1** | | | | |
|---|---|---|---|---|
| **Variables\bandwidth m** | **m:18** | **m:20** | **m:23** | **m:28** |
| **Andal** | 2.63 | 2.31 | 2.44 | 1.47 |
| **Ara** | 1.58 | 0.94 | 0.879 | 0.24 |
| **Ast** | 0.81 | 1.62 | 1.20 | 0.35 |
| **Bal** | 1.30 | 1.18 | 0.622 | 0.06 |
| **Can** | 1.06 | 1.64 | 1.23 | 0.62 |
| **Cant** | 1.03 | 1.82 | 1.75 | 0.85 |
| **Cat** | 1.57 | 1.46 | 1.32 | 0.86 |
| **Clm** | 2.32 | 1.85 | 1.88 | 1.29 |
| **Cyl** | 0.48 | 0.28 | 0.242 | 0.16 |
| **Ext** | 0.91 | 0.317 | 0.111 | 0.046 |
| **Gal** | 1.51 | 1.45 | 1.42 | 1.32 |
| **Mad** | 1.69 | 2.55 | 2.26 | 1.56 |
| **Mur** | CT | CT | CT | CT |
| **Nav** | 1.58 | 1.66 | 1.52 | 0.73 |
| **Pv** | 0.39 | 1.30 | 1.21 | 0.61 |
| **Rio** | 0.12 | 0.019 | 0.00 | 0.24 |
| **Val** | 1.49 | 1.89 | 1.79 | 1.19 |

*Note: The series are the growth rate per capita.*

Table 6. Cointegration rank test, step XVI

| **Step 16** | | | | |
|---|---|---|---|---|
| **Variables\bandwidth m** | **m:18** | **m:20** | **m:23** | **m:28** |
| **Andal** | CT | CT | CT | CT |
| **Ara** | CT | CT | CT | CT |
| **Ast** | CT | CT | 0.092 | 0.0014 |
| **Bal** | CT | CT | CT | CT |
| **Can** | CT | CT | CT | CT |
| **Cant** | CT | CT | CT | CT |
| **Cat** | CT | CT | CT | CT |
| **Clm** | 0.0019 | CT | CT | CT |
| **Cyl** | CT | CT | CT | CT |
| **Ext** | CT | CT | CT | CT |
| **Gal** | CT | CT | CT | CT |
| **Mad** | CT | 0.06 | CT | CT |
| **Mur** | CT | CT | CT | CT |
| **Nav** | CT | CT | CT | CT |
| **Pv** | CT | CT | CT | CT |
| **Rio** | CT | CT | CT | CT |
| **Val** | CT | CT | CT | CT |

### 3.3. Nielsen's (2010) cointegration results

Unlike the cointegration test proposed by Robinson (2008), the procedure in Nielsen (2010) begins with $H_0: r = r_0$ versus $H_1: r > r_0$ and the testing sequence ends when the null hypothesis is not rejected. The variance ratio rank test is based on the statistic $\Lambda_{p,r}$, defined in the above section, for p=1 to 17 and applied to the non-stationary raw series. Denoting $CV_{\xi,p-r}(d, d_1)$ the critical values, if $\Lambda_{p,r}(d_1) > CV_{\xi,p-r}(d, d_1)$ the null hypothesis is rejected, and we can continue to the next step. The results of this variance ratio test applied in our sample are shown in Table 7. Panel B shows the corresponding variance ratio test for the case with constant and trend regarding the behaviour of the raw series in Figure 2.

Comparing the statistics with the critical values with $\delta_t = [1, t]'$ in Table 2, with a significance level of 5%, we have $\Lambda_{17,0}(d_1) < CV_{0.05,17}(d, d_1) = 34.96 < 35.84$, concluding that the null hypothesis $r = 0$ cannot be rejected. Note that for a significance level of 10%, the results are the same (we cannot reject the null hypothesis). This result indicates the presence of no cointegration among the per capita gross domestic product of the autonomous communities in Spain, reinforcing the lack of convergence found with the semiparametric test developed by Robinson (2008).

Table 7. Variance ratio cointegration rank test

| Panel B: Vartest $\Lambda_{p,r}(0.1)$ | | | | | | | | | | | | | | | | |
|---|---|---|---|---|---|---|---|---|---|---|---|---|---|---|---|---|
| | | | | | | | $p - r$ | | | | | | | | | |
| 1 | 2 | 3 | 4 | 5 | 6 | 7 | 8 | 9 | 10 | 11 | 12 | 13 | 14 | 15 | 16 | 17 |
| 1.50 | 3.10 | 4.82 | 6.59 | 8.46 | 10.38 | 12.35 | 14.38 | 16.47 | 18.66 | 20.88 | 23.14 | 25.43 | 27.75 | 30.09 | 32.50 | 34.96 |

Note: Panel B reports the variance ratio cointegration test statistic $\Lambda_{p,r}(d_1)$

## 4. Conclusions

In this work we have tested for the existence of convergence of real per capita GDP in 17 autonomous communities of Spain from 1955 to 2020, applying novel time-series methodologies in a fractional integration and cointegration context. Our analyses yielded no evidence of long-run equilibrium relationships, thus ruling out the possibility of convergence according to the definitions of Bernard and Durlauf (1995). Since cointegration is a necessary condition for output convergence, economic convergence is rejected in this sample. The results are robust, confirmed with different techniques used to test for cointegration and are also consistent with a cross sectional analysis based on sigma-convergence.

Our empirical findings provide us with an important insight into this framework. There is no overall economic convergence among the autonomous communities, and no convergent subgroup of regions has been identified. The complete absence of convergence among the 17 autonomous communities might be explained by the fact that the economic differences and performance between them are significant and persistent, and as such, the convergence process has not been able to offset them within the analyzed time frame. This may be due to the existence of heterogeneous production functions, different levels of human capital, or regional differences in the quality of institutions, among other factors, that affect the growth rates of the regions differently. In addition, the existence of sector-specific shocks and productivity spillovers may also influence the speed of convergence. Moreover, the convergence process may be affected by various economic and policy factors, such as trade openness, fiscal transfers, and labour mobility, which may have different effects across regions and over time.

This underlies the economic disparities between regions, potentially leading to non-convergence. This absence of convergence and lack of general long run equilibrium should be taken into account by the Spanish authorities when designing economic policies, focusing on regional specific better that general policies.